\documentclass{PoS}

\usepackage{subfigure}
\usepackage{amssymb,amsmath}

\title{Scaling behavior and sea quark dependence of pion spectrum
with HYP-smeared staggered fermions}

\ShortTitle{Scaling behavior and sea quark dependence of pion spectrum}

\author{David Adams, \speaker{Taegil Bae}, Hyung-Jin Kim,
Jongjeong Kim, Kwangwoo Kim, Boram Yoon, and Weonjong Lee \\
Frontier Physics Research Division and
Center for Theoretical Physics, \\
Department of Physics and Astronomy, 
Seoul National University, Seoul, 151-747, South Korea \\ 
E-mail: \email{wlee@phya.snu.ac.kr}}

\author{Chulwoo Jung \\ 
Physics Department, Brookhaven National Laboratory, 
Upton, NY11973, USA \\
E-mail: \email{chulwoo@bnl.gov}}

\author{Stephen R.~Sharpe \\ 
Department of Physics, University of Washington, Seattle,
WA 98195-1560, USA \\ 
E-mail: \email{sharpe@phys.washington.edu}}

\abstract{ 
We study the pion spectrum 
(and in particular taste-symmetry breaking
within it) using HYP-smeared valence staggered
fermions on the coarse and fine MILC lattices (which have
asqtad staggered sea quarks). We focus on the dependence
on lattice spacing and sea-quark mass. We also update our results
on source dependence. Our main conclusion is that on the MILC
fine lattices the appropriate power-counting for SU(3) 
staggered chiral perturbation theory may have 
discretization errors entering at next-to-leading order rather
than at leading-order.
}

\FullConference{The XXVI International Symposium on Lattice Field Theory \\
                July 14 - 19, 2008\\
                Williamsburg, Virginia, USA}

\begin{document}
\section{Introduction}
\label{sec:intro}
In this note we give an update on our results 
(most recently presented in Ref.~\cite{ref:wlee:0})
for the pion spectrum using valence 
HYP-smeared staggered fermions. These results
are needed as input into our parallel calculation of $B_K$
(whose status is described in Ref.~\cite{ref:wlee:lat08}),
and more generally provide information on the systematic errors
that are likely to found with improved staggered actions
(including the
``highly improved staggered quark'' [HISQ] action~\cite{ref:HISQ}).

Smearing the gauge links in the fermion action is a widely used 
method to reduce discretization and perturbative errors. We use the
unimproved staggered action with HYP-smeared~\cite{ref:hyp:1} links.
One conclusion of Ref.~\cite{ref:wlee:0} was
that this smearing reduced the dominant discretization 
error---taste-symmetry breaking in the pion spectrum---by
a factor $2.5-3$ compared to quarks improved with
the asqtad action~\cite{ref:asqtad:1}.
The resulting taste-breaking was thus reduced to
the same level as with the HISQ action.
Thus, for light quarks, where the additional improvements in
the HISQ action are less important, HYP-smeared quarks represent
a simpler alternative to the HISQ action.

The reduction in taste-breaking was studied in Ref.~\cite{ref:wlee:0}
on the coarse MILC lattices (with $a\approx 0.125\,$fm).
For these lattices it was found that, even after HYP-smearing,
the size of $O(a^2)$ errors was comparable, for the lightest quark
masses, to the $O(p^2)$ terms in chiral perturbation theory ($\chi$PT).
Thus the standard $a^2\sim  p^2$ power-counting of
staggered chiral perturbation theory (S$\chi$PT) must still be used.
It was suggested, however, that on the fine MILC lattices
(with $a\approx 0.09\,$fm) taste-breaking would be reduced to
the extent that the appropriate power-counting would
be $a^2\sim p^4$, thus simplifying the required fitting.
Our new results allow us to study this point.

Another observation of Ref.~\cite{ref:wlee:0} was that
$O(a^2 p^2)$ effects were significantly smaller than expected. 
% (for both HYP-smeared and asqtad quarks). 
With improved statistics we can now quantify this statement.

Our final new result concerns the %(lack of)
dependence of the pion spectrum on the sea-quark mass. 

\section{Scaling behavior of pion spectrum}
\label{sec:scaling}
\begin{figure}[t!]
  \centering
  \subfigure[MILC coarse lattices ($a=0.12$fm)]{
    \includegraphics[width=.45\textwidth]
                    {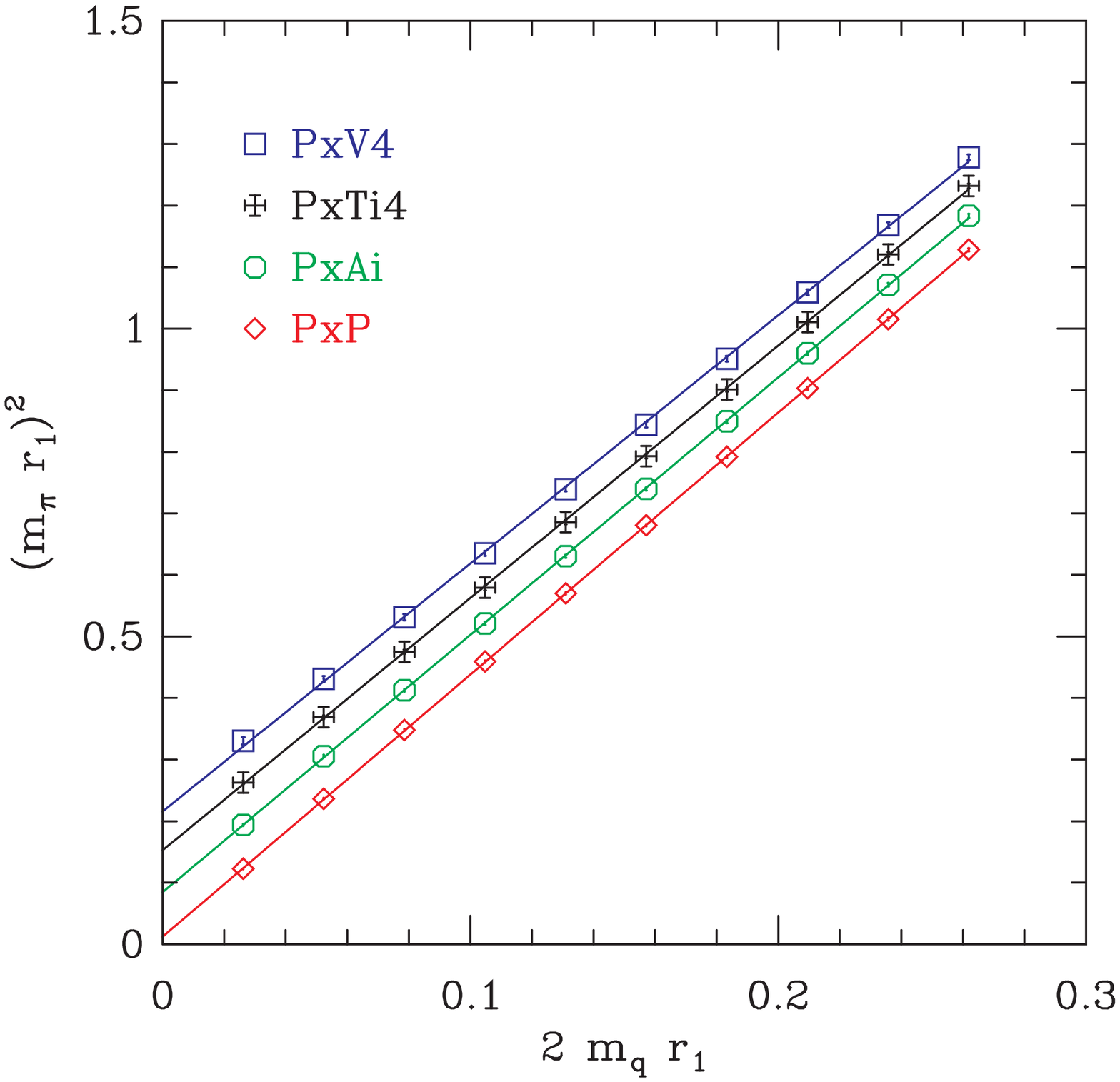}}
  \subfigure[MILC fine lattices ($a=0.09$fm)]{
    \includegraphics[width=.45\textwidth]
                    {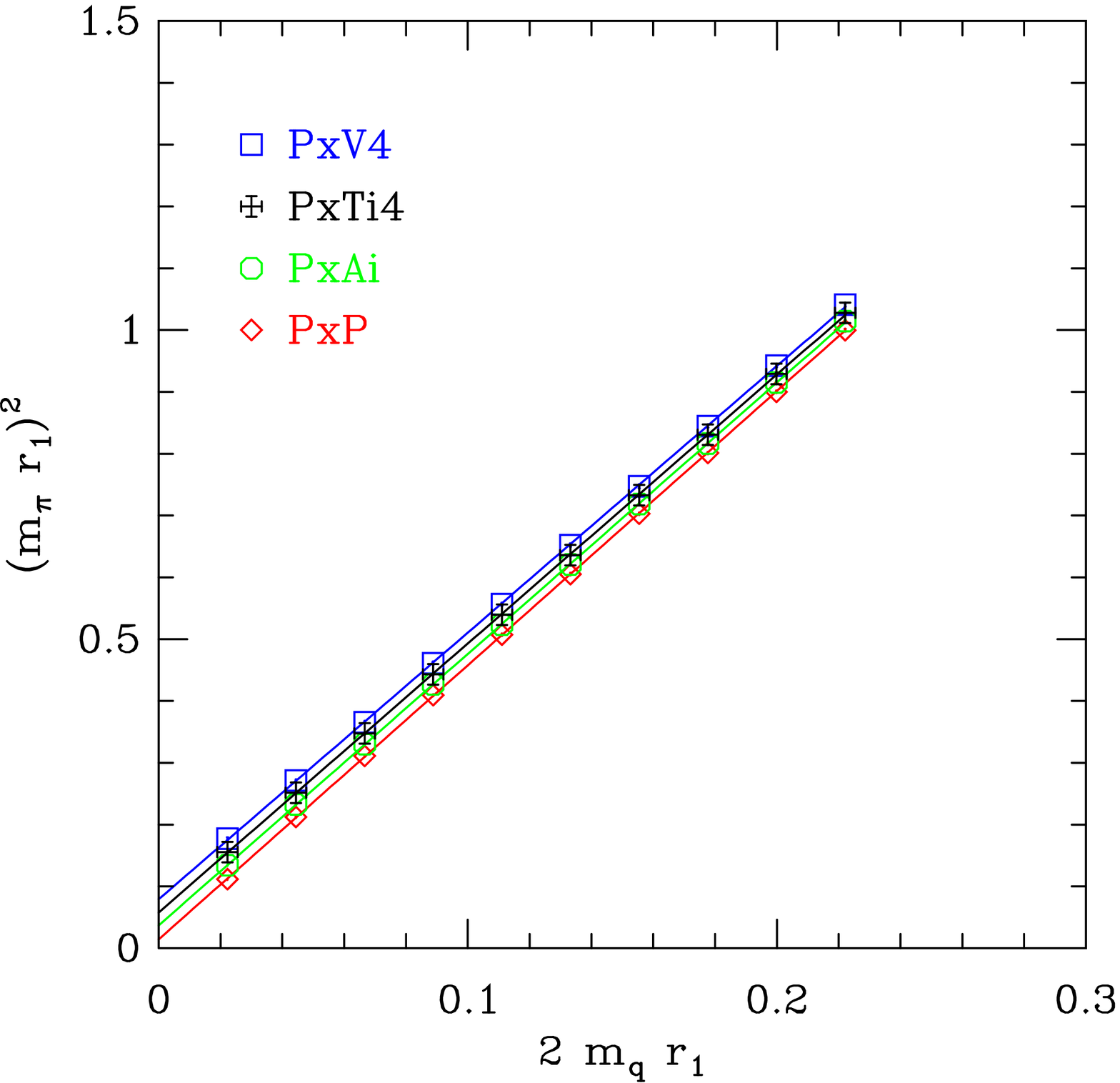}}
\caption{$m_\pi^2$ vs. $2m_q$: the data is obtained using HYP-smeared
  staggered fermions with cubic wall sources and Golterman sink
  operators \cite{ref:wlee:0}. The legend indicates the tastes.}
\label{fig:scaling:1}
\end{figure}
%%%%%%%%%%%%%%%%%%%%%%%%%%%%%%%%%
%\begin{figure}[t!]
%  \centering
%  \subfigure[Coarse lattices]{
%    \includegraphics[width=.48\textwidth]
%                    {mpisq_mq_r1_mixed_gol_ss_cw_MILC_2064f21b676m010m050}}
%  \subfigure[Fine lattices]{
%    \includegraphics[width=.48\textwidth]
%                    {mpisq_mq_r1_mixed_gol_ss_cw_MILC_2896f21b709m0062m031}}
%\end{figure}
%%%%%%%%%%%%%%%%%%%%%%%%%%%%%%%%%%
%
%
\begin{table}[b]
  \begin{center}
    \begin{tabular}{c|c|c}
      \hline
      gauge action & \multicolumn{2}{c}{1-loop tadpole-improved Symanzik 
        gluon action} \\
      sea quarks & \multicolumn{2}{c}{$N_f=2+1$ Asqtad staggered fermions}\\
      valence quarks  & \multicolumn{2}{c}{HYP staggered fermions} \\
      \hline \hline 
      parameters & MILC fine lattices & MILC coarse lattices \\
      \hline
      $a$ & $0.09\textrm{fm}$ & $0.12\textrm{fm}$\\
      geometry & $28^{3}\times96$ & $20^3\times64$ \\
      \# of confs & $995$ & $671$ \\
      sea quark masses & $am_{l}=0.0062,$  & $am_l=0.01,$ \\
                       & $am_{s}=0.031$    & $am_s=0.05$ \\
      valence quark masses & $0.003, 0.006, \dotsc, 0.030$ & 
      $0.005, 0.01, \dotsc, 0.05$ \\
      \hline
    \end{tabular}
  \end{center}
\caption{Parameters for the numerical study on scaling violations.}
\label{tab:para:1}
\end{table}
In Fig.~\ref{fig:scaling:1}, we show $m_\pi^2$ as a function of quark
mass for both coarse and fine MILC lattices with 
sea quark mass-ratio of $m_\ell/m_s=1/5$.
Parameters for the numerical study are summarized in Table
\ref{tab:para:1}---we note that the lightest valence quarks
have a mass of $\approx m_s^{\rm phys}/10$.
We show only results with degenerate valence quarks, and
display a linear fit.
For clarity,
we show only the four tastes with the smallest errors
(those created by pseudoscalar operators
which are local in time~\cite{ref:wlee:0}).

\begin{figure}[t!]
\begin{center}
  \includegraphics[width=0.55\textwidth]{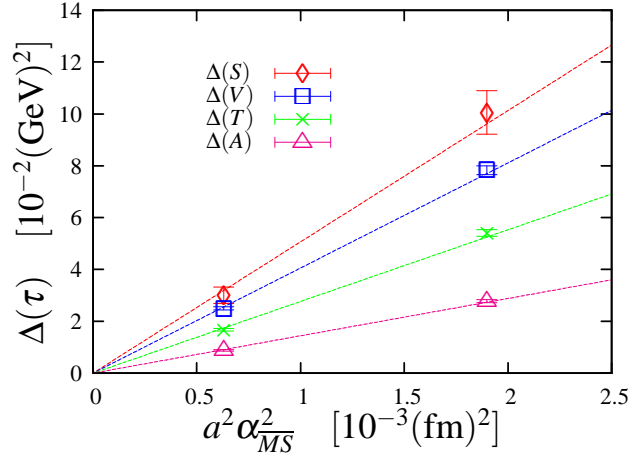}
\end{center}
\caption{$\Delta(T)$ vs. $a^2 \alpha_{\overline{\rm MS}}^2$:
  $\alpha_{\overline{\rm MS}}^2$ is obtained at $\mu=1/a$ using
  four-loop running.}
\label{fig:scaling:2}
\end{figure}

Taste-breaking is substantially reduced on the fine lattices.
To make this quantitative, we tabulate results for
the splitting in the chiral limit
\begin{equation}
\Delta(T) = \lim_{m_q \rightarrow 0} [m_\pi^2(T) - m_\pi^2(\xi_5)]\,,
\end{equation}
with $T$ the taste, in Table \ref{tab:delta:1}.
Our results are consistent with the $SO(4)$ symmetry expected
at $O(a^2)$~\cite{ref:wlee:1,ref:bernard:1}, and we quote the
result for the taste with the smallest error.
The ratio $\Delta(T,{\rm fine})/\Delta(T,{\rm coarse})$
is about $0.3$ for all tastes.
\begin{table}[b]
  \begin{center}
    \begin{tabular}{c|l|l|c}
      \hline
      & \multicolumn{2}{c|}{$\Delta(T)$ [$({\rm GeV})^2$]} & 
      \\ \cline{2-3}
      \raisebox{1.5ex}[0cm][0cm]{Taste ($T$)}
      & $a = 0.125{\rm fm}$ & $a = 0.09 {\rm fm}$
      & \raisebox{1.5ex}[0cm][0cm]{$\Delta({\rm Fine})/\Delta({\rm Coarse})$}
      \\ \hline
      $\xi_{5\mu}$   & 0.0278(6)      & 0.0087(3)       & 0.314(12) \\
      $\xi_{\mu\nu}$ & 0.0540(13)     & 0.0168(4)       & 0.310(11) \\
      $\xi_{\mu}$    & 0.0783(17)     & 0.0250(6)       & 0.319(11) \\
      $I$            & 0.1005(84)     & 0.0300(31)      & 0.299(40) \\
      \hline
    \end{tabular}
  \end{center}
  \caption{$\Delta(T)$ in physical units on coarse and fine MILC lattices.}
  \label{tab:delta:1}
\end{table}
%
% a^2 alpha terms are removed by improvement.
%  expect scaling a^2 alpha^2
% the meaning of the splitting = intercept
%
For HYP-smeared staggered fermions,
taste-symmetry breaking 
is of order ${\cal O}(a^2 \alpha_s^2)$.\footnote{%
%
% $\alpha_s = g^2/(4\pi)$ is the gauge coupling constant in QCD. 
Smearing removes
${\cal O}(a^2 \alpha_s)$ taste symmetry breaking terms.}
Our results are consistent with this expectation,
as shown in Fig.~\ref{fig:scaling:2}.
%\footnote{%
%{\bf Note to Weonjong: presumably $\alpha_s$ is obtained starting
%from $m_Z$? Perhaps we don't need to get into these details.
%Alternatively, we could use the ratio of coupling constants used
%by MILC??? Let's discuss}
%}
%

In light of these results,
what is the appropriate power-counting 
for S$\chi$PT 
% staggered chiral perturbation theory 
on the fine lattices?
When considering kaon properties (e.g. $B_K$), there are loops
containing valence $d \bar d$ pions of all tastes.
For our lightest valence quark, the ratio of $O(a^2)$
to $O(p^2)$ contributions is characterized by
\begin{equation}
(m_\pi^2(\xi_{\mu\nu})-m_\pi^2(\xi_5))/m_\pi^2(\xi_{\mu\nu})
\equiv
\Delta(\xi_{\mu\nu})/m_\pi^2(\xi_{\mu\nu}) 
\approx 0.25 \,.
\end{equation}
(We use tensor taste as it is the most numerous and
its mass lies near the average of the multiplet.)
Since $\chi$PT for kaons is characterized by
an expansion parameter of $(m_K^{\rm phys}/1\ {\rm GeV})^2 \approx 25\%$, 
it may be appropriate to use the power-counting $O(a^2) \sim O(p^4)$,
i.e. to treat discretization errors as of next-to-leading order (NLO).
This will not, however, be appropriate for calculations
of $f_\pi$ using $SU(2)$ chiral perturbation theory, where the
expansion parameter is smaller.

Further information on the power counting can be obtained by
comparing the slopes of the linear fits, differences between
which are of $O(a^2 p^2)$. Our results for these slopes
for both lattices and all tastes
are given in Table~\ref{tab:slope:1}.
\begin{table}[bt]
  \begin{center}
    \begin{tabular}{|c|c|c|}
      \hline
      & \multicolumn{2}{c|}{slope($T$)} 
      \\ \cline{2-3}
      \raisebox{1.5ex}[0cm][0cm]{Taste ($T$)}
      & $a = 0.125{\rm fm}$ & $a = 0.09 {\rm fm}$
      \\ \hline
      $\xi_{5}$    & 4.258(11) & 4.436(12)        \\
      $\xi_{i5}$   & 4.182(14) & 4.391(12)        \\
      $\xi_{i4}$   & 4.100(20) & 4.352(13)        \\
      $\xi_{4}$    & 4.031(23) & 4.314(15)        \\
      \hline
      $\xi_{45}$   & 4.180(33) & 4.414(33)        \\
      $\xi_{ij}$   & 4.115(42) & 4.373(35)        \\
      $\xi_{i}$    & 4.052(53) & 4.340(37)        \\
      $I$          & 3.952(73) & 4.312(40)       \\
      \hline
    \end{tabular}
  \end{center}
  \caption{Slopes, $c_2$, of the
    linear fits  $(r_1 m_\pi(T))^2 = c_1 + c_2
    (2r_1 m_q)$.}
  \label{tab:slope:1}
\end{table}
We observe that there are significant splittings between
slopes for different tastes (the differences are
more significant than the errors suggest because of correlations),
but that these splittings are small.
For example, the splitting between slopes for
tastes $\xi_5$ and $\xi_{i4}$ are $\approx 4\%$ and $2\%$
on coarse and fine lattices, respectively.
On the fine lattices, this is consistent the power-counting
$O(a^2)\sim O(p^4)$, in which case the splittings between
slopes are a NNLO effect.

We also note that $SO(4)$ {\em breaking} in the slopes
is much smaller than the splittings allowed by $SO(4)$,
although both are effects of the same order,
$O(a^2 p^2)$~\cite{ref:SVdW}.
This remains to be understood.

\begin{figure}[t!]
  \centering
  \subfigure[Scaling violation]{
    \includegraphics[width=.45\textwidth]
                    {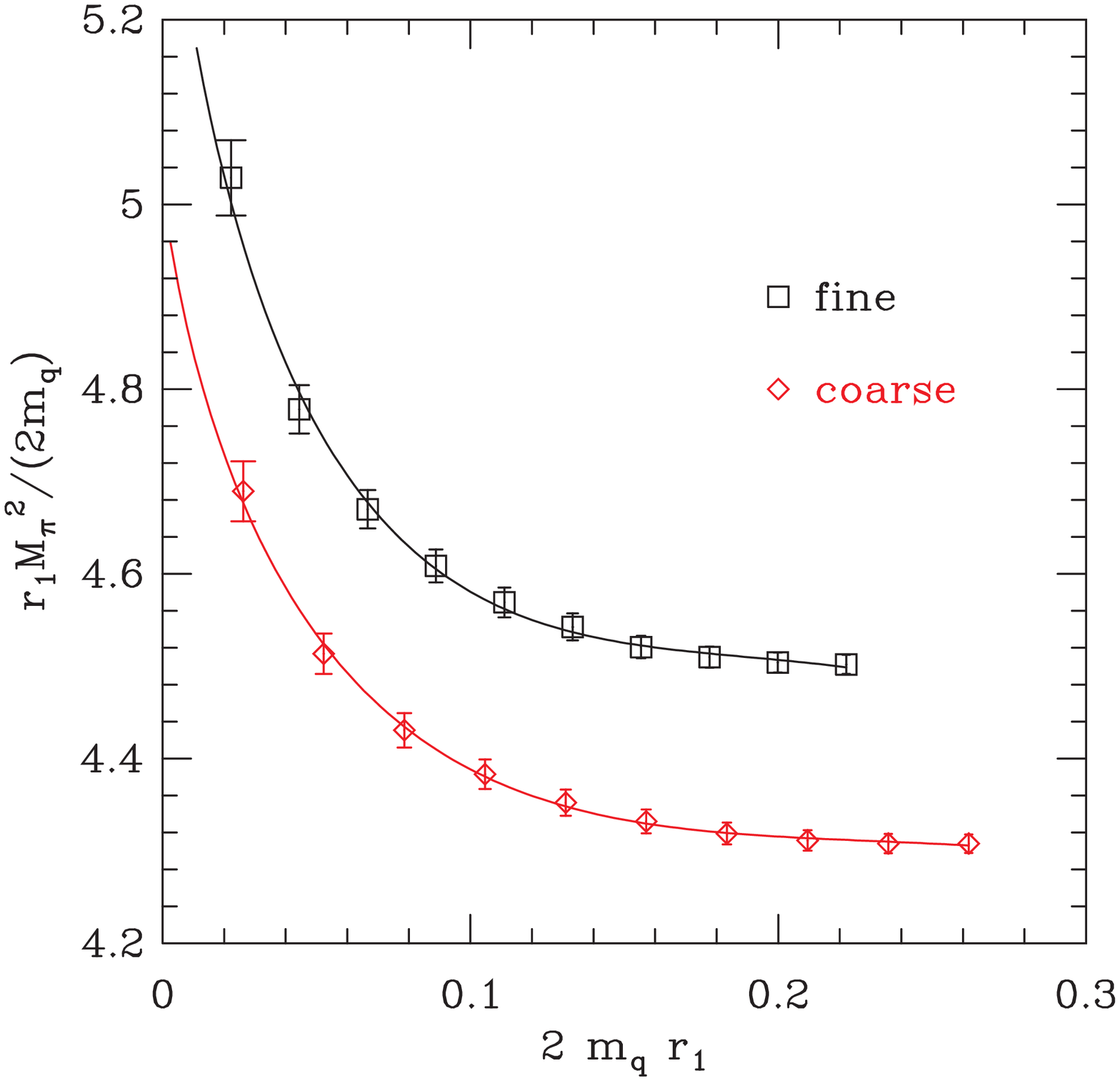}}
  \subfigure[Sea quark dependence]{
    \includegraphics[width=.45\textwidth]
                    {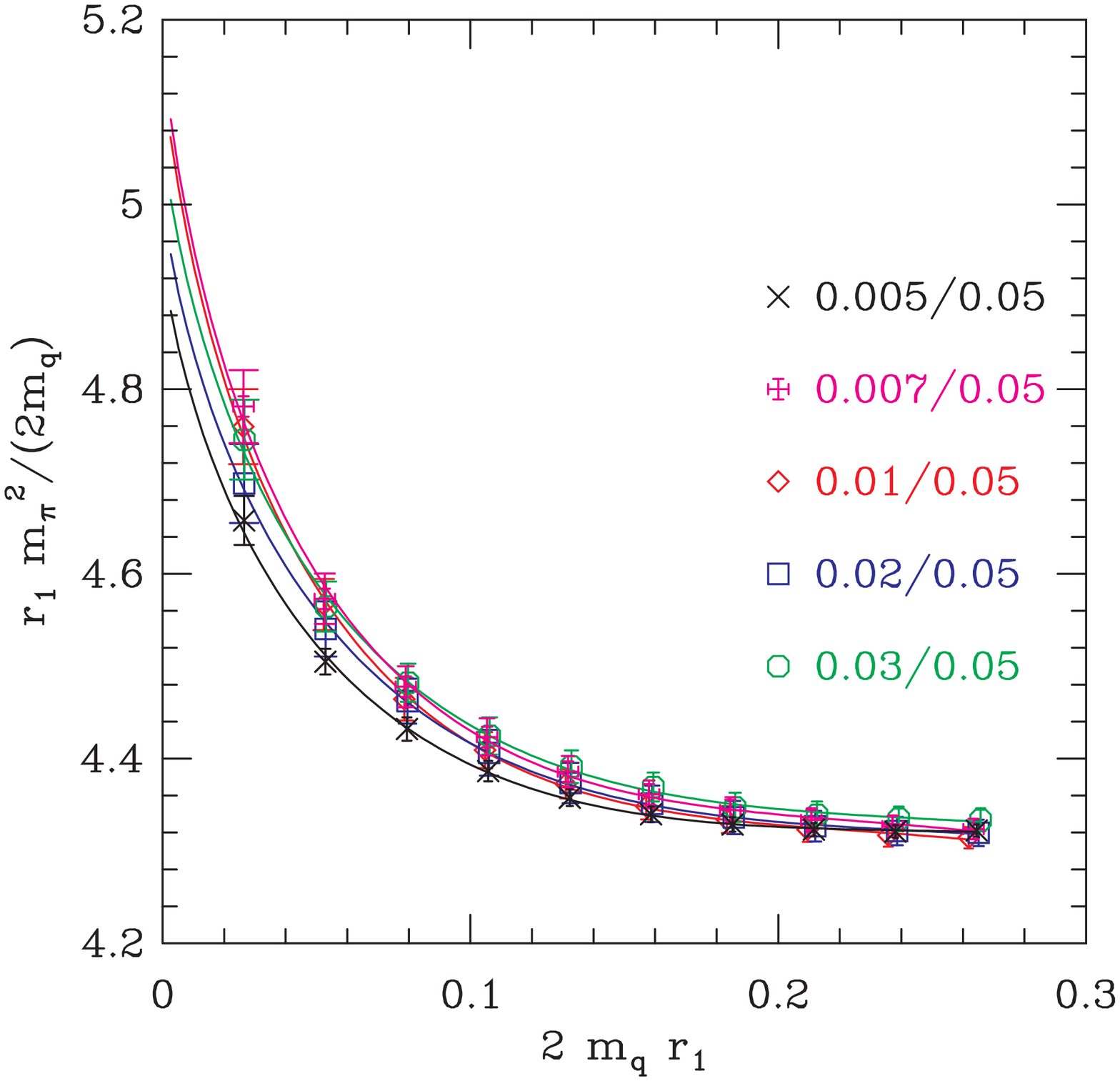}}
\caption{$r_1 m_\pi^2 / (2m_q)$ vs. $2 r_1 m_q$ for the Goldstone pion.}
\label{fig:scaling:3}
\end{figure}

Finally, we note that the difference between the slopes
for the two lattice spacings, which is roughly 5\%,
is of the expected size for a discretization effect
(i.e. $(a\Lambda_{\rm QCD})^2$).
This is also seen in Fig.~\ref{fig:scaling:3}(a).
We note that this difference is 
comparable to the taste splittings
on the coarse lattices, indicating that the latter are not
significantly enhanced over taste-symmetric discretization errors.

\section{Sea quark dependence of pion masses}
\begin{figure}[t!]
  \centering
    \includegraphics[width=.43\textwidth]
                    {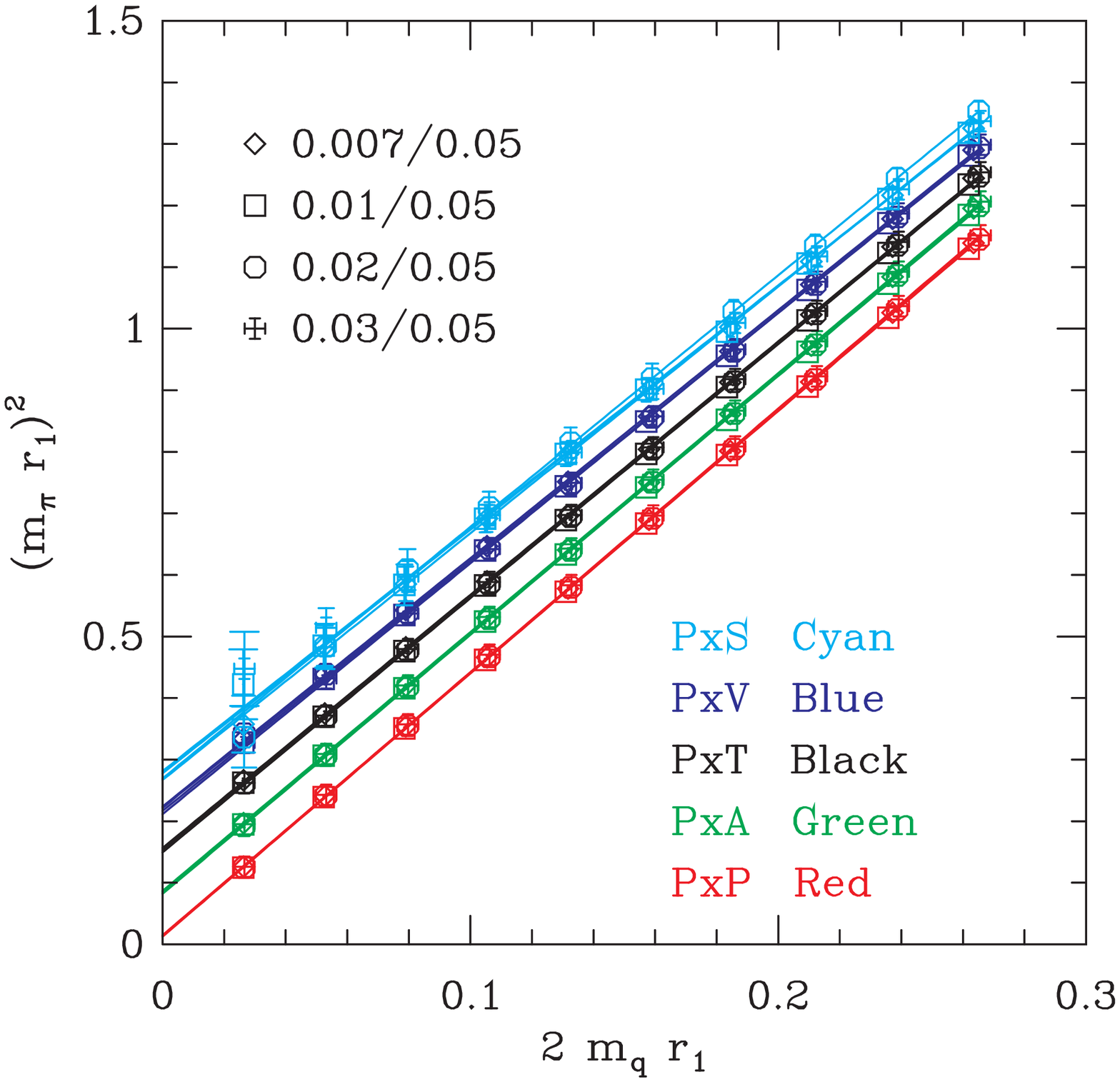}
    \includegraphics[width=.48\textwidth]
                    {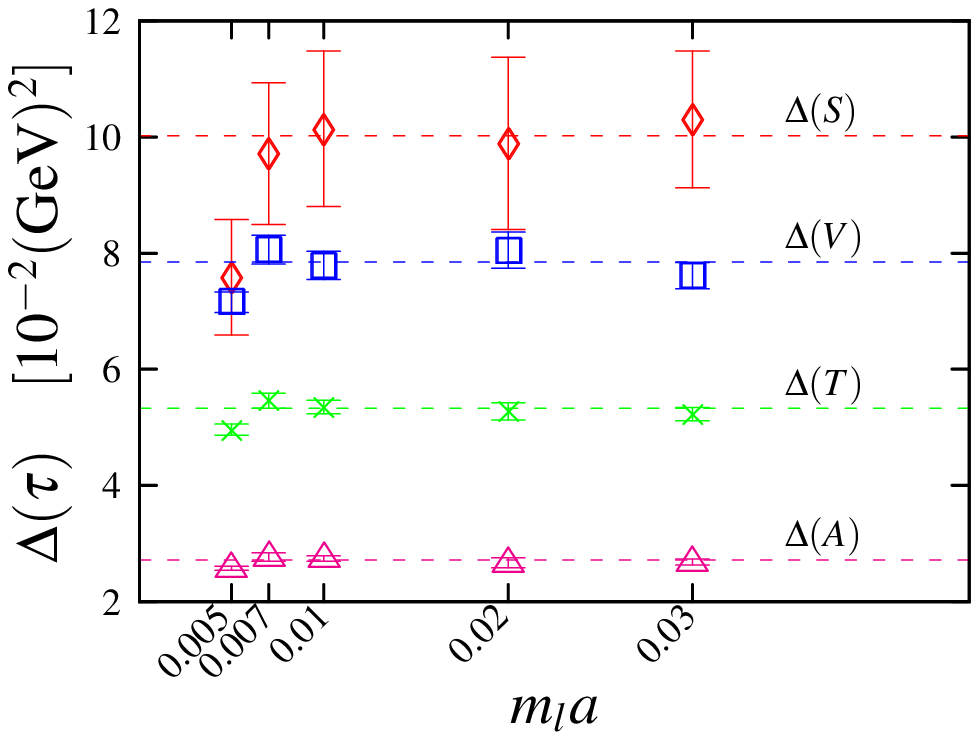}
\caption{$(m_\pi r_1)^2$ vs. $2 m_q r_1$(left) and 
  $\Delta(T)$ vs. $a m_l({\rm sea})$(right)}
\label{fig:sea:1}
\end{figure}

In Table \ref{tab:sea:1}, we list the MILC coarse-lattice ensembles
we use to study the dependence of the pion spectrum on sea quark masses.
%
%Basically, we fix the strange quark mass and vary the light quark mass.
%
%The $m_{\pi,{\rm sea}} L$ changes from 7.6 to 3.8.
%
\begin{table}[h]
  \begin{center}
    \begin{tabular}{c|c|c|c}
      \hline
      Geometry & $am_l$(sea) & $am_s$(sea) & $m_{\pi,{\rm sea}} L$ \\
      \hline
      $24^3\times 64$ & $0.005$ & $0.05$ & $3.83$ \\
      $20^3\times 64$ & $0.007$ & $0.05$ & $3.78$ \\
      $20^3\times 64$ & $0.010$ & $0.05$ & $4.49$ \\
      $20^3\times 64$ & $0.020$ & $0.05$ & $6.23$ \\
      $20^3\times 64$ & $0.030$ & $0.05$ & $7.56$ \\
      \hline
    \end{tabular}
  \end{center}
\caption{Parameters for numerical study of sea quark mass dependence
on coarse lattices.}
\label{tab:sea:1}
\end{table}
We find no significant dependence within our errors.
This is shown for the Goldstone pion in Fig.~\ref{fig:scaling:3}(b),
and for all tastes in Fig.~\ref{fig:sea:1}.
Note that in Fig.~\ref{fig:sea:1},
different colors represent different tastes while different
symbols corresponds to different light sea-quark masses.

The righthand plot in Fig.~\ref{fig:sea:1} shows $\Delta(T)$ as a
function of the light sea quark mass.
There is a 2 $\sigma$ drop for 
$am_{l,{\rm sea}} = 0.005$, but only improved
statistics will allow us to determine if this is
significant.

The independence of the pion spectrum on the sea-quark mass
is consistent with the results of the MILC collaboration with
asqtad quarks~\cite{ref:milc:1}. 
%
% Our data shows a range of about 0.03, and no particular ordering
%
% Milc has a range of about 0.04, but the overall scale is 1.5x higher
% They also have no particular order.
% The only exception is the 0.03/0.05 data which lies lower for MILC
%
We note, however, that other
quantities, e.g. $f_\pi$ and $f_K$, do show some dependence
on $m_{\rm sea}$.

\section{Efficacy of different sources}
%
%
%%%%%%%%%%%%%%%%%%%%%%%%%%%%%%%%%%%%
%\begin{figure}[t!] \centering
%  \subfigure[Cubic wall \lbrack CW\rbrack]{
%    \includegraphics[width=.48\textwidth]
%                    {mpisq_mq_r1_mixed_gol_cw_MILC_2064f21b676m010m050}}
%  \subfigure[Cubic U(1) \lbrack CU(1)\rbrack]{
%    \includegraphics[width=.48\textwidth]
%                    {mpisq_mq_r1_mixed_gol_cu1_MILC_2064f21b676m010m050}}
%\end{figure}
%
%
%
%\begin{figure}[t!] \centering
%  \subfigure[Cubic wall (CW)]{
%    \includegraphics[width=.48\textwidth]
%                    {mpisq_mq_r1_mixed_gol_ss_cw_MILC_2064f21b676m010m050}}
%  \subfigure[Cubic U(1) (CU(1))]{
%    \includegraphics[width=.48\textwidth]
%                    {mpisq_mq_r1_mixed_gol_ss_cu1_MILC_2064f21b676m010m050}}
%\end{figure}
%%%%%%%%%%%%%%%%%%%%%%%%%%%%%%%
%
Finally, we have updated our results on the 
relative efficacy of our two sources:
``cubic wall'' and ``cubic U(1)'' sources.
(For their definitions see Ref.~\cite{ref:wlee:0}.)
Both sources project onto
specific irreps of the cubic group, but we do not know
{\em a-priori} what the resulting signal/noise ratio will be.
Here we extend the study of Ref.~\cite{ref:wlee:0} by
comparing the sources on
the MILC coarse lattice ensemble with $am_l=0.01$ and $am_s=0.05$.

\begin{figure}[t!]
  \parbox{4cm}{
  \begin{tabular}{c|c|c}
    \hline
    & \multicolumn{2}{c}{$\Delta(T)$ [$({\rm GeV})^2$]}\\ \cline{2-3}
    \raisebox{1.5ex}[0cm][0cm]{Taste} & Cubic wall [CW] & Cubic U(1) [CU1]\\
    \hline
    $\xi_i\xi_5$ & 0.0278(6)      & 0.0274(5) \\
    $\xi_4\xi_i$ & 0.0540(13)     & 0.0535(11) \\
    $\xi_4$      & 0.0783(17)     & 0.0779(24) \\
    \hline
    $\xi_4\xi_5$ & 0.0253(33)     & 0.0240(49) \\
    $\xi_i\xi_j$ & 0.0500(43)     & 0.0486(61)  \\
    $\xi_i$      & 0.0740(57)     & 0.0773(101) \\
    $I$          & 0.1005(84)     & 0.1014(134) \\
    \hline
  \end{tabular} }
\hspace*{3.5cm}
  \parbox{4cm}{
  \includegraphics[width=.48\textwidth,clip=true]{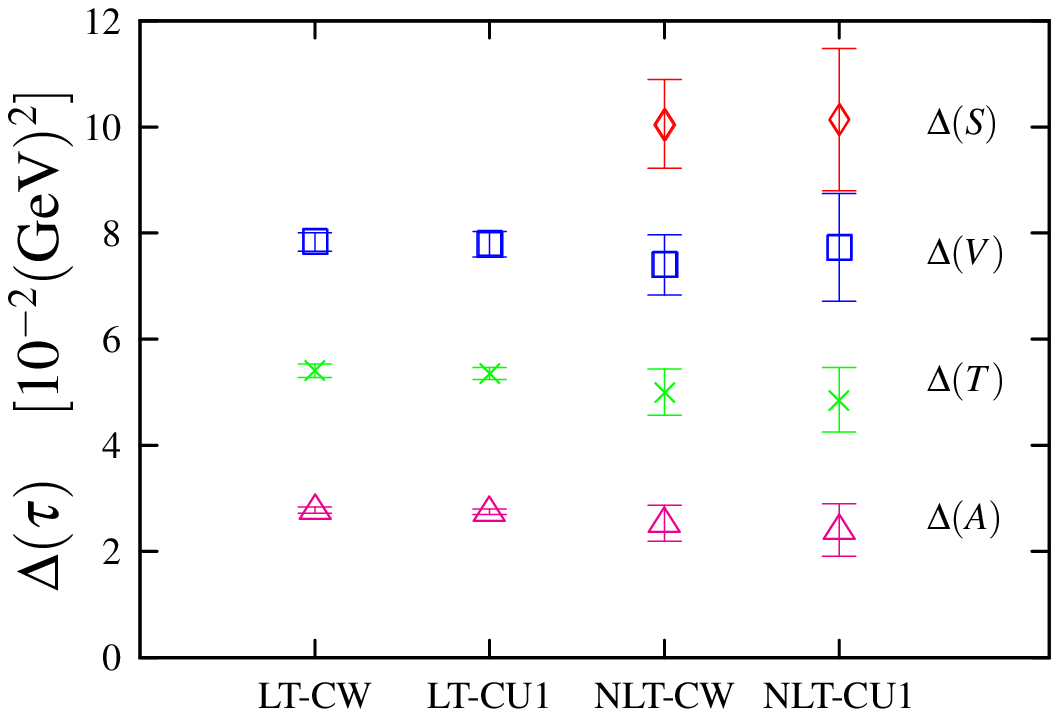}}
\caption{$\Delta(T)$ for cubic wall and cubic U(1) sources.}
\label{fig:cw-cu1:1}
\end{figure}
In Fig.~\ref{fig:cw-cu1:1}, we present $\Delta(T)$ for cubic wall
(CW) and U(1) sources (CU1).
In this plot, LT represents ``local-in-time'' tastes ($\{\xi_i\xi_5,
\xi_4\xi_i,\xi_4\}$) and NLT represents ``non-local-in-time'' tastes
($\{\xi_4\xi_5, \xi_i\xi_j, \xi_i, I \}$).
Thus, for example, $\xi_i\xi_5$ and $\xi_4\xi_5$ form two different
irreps of $SW_4$ but belong to the same irrep of $SO(4)$.
S$\chi$PT predicts that the pion spectrum
should respect the $SO(4)$ symmetry in the chiral limit,
both at LO~\cite{ref:wlee:1, ref:bernard:1}
and at NLO~\cite{ref:SVdW}.
In Fig.~\ref{fig:cw-cu1:1}, we notice that $SO(4)$ symmetry is well
respected in $\Delta(T)$ since the data for LT tastes and NLT
tastes are consistent with each other within statistical uncertainty.

In the case of LT tastes, the results from both sources
have comparable errors.
%
%Hence, both sources are equally effective to select a specific pion
%state of LT tastes.
%
However, for the NLT tastes, CW sources are clearly preferred,
with errors being smaller by about a factor of 2.
This is a more significant difference than we observed in
Ref.~\cite{ref:wlee:0} on quenched lattices.

\section{Conclusions}

We have observed the expected reduction in taste-symmetry breaking
by a factor of $\approx 3$ when passing from the coarse to the fine
MILC lattices. This implies that, for $SU(3)$ staggered
chiral perturbation theory, using HYP-smeared valence
quarks probably allows one to treat discretization
errors as a NLO effect. Whether this is really the case can only
be determined by detailed
fitting, such as that we are undertaking for $B_K$.
It will also be interesting to do NLO fits of our pion-spectrum data
itself, but this requires NLO expressions for pions of all tastes,
which are in progress but not yet completed.

Our finding that the pion spectrum is almost independent of the sea-quark 
mass, which is consistent with the MILC asqtad results, simplifies our
calculation of $B_K$, since we need not recalculate the entire spectrum
for each MILC fine lattice (although we will check this by calculating
the Goldstone taste).

\section{Acknowledgment}
\label{sec:ack}
C.~Jung is supported by the US DOE under contract DE-AC02-98CH10886.
The research of W.~Lee is supported in part by the KICOS grant
K20711000014-07A0100-01410, by the KRF grants (KRF-2007-313-C00146 and
KRF-2008-314-C00062), by the BK21 program, and by the US DOE SciDAC-2
program.
The work of S.~Sharpe is supported in part by the US DOE grant
no. DE-FG02-96ER40956, and by the US DOE SciDAC-2 program.

%%%%%%%%
% EDIT
%%%%%%%%

%
%
%


\begin{thebibliography}{99}
%
\bibitem{ref:wlee:0} T.~Bae, {\em et al.},
  Phys.~Rev.~D{\bf 77} (2008) 094508, [{arXiv:0801.3000}].
%
\bibitem{ref:wlee:lat08} T.~Bae, {\em et al},
  PoS (LATTICE 2008) 275, [{arXiv:0809.1220}].
%
\bibitem{ref:HISQ}   E.~Follana {\it et al.}  [HPQCD Collaboration],
  %``Highly improved staggered quarks on the lattice, with applications to
  %charm physics,''
  Phys.\ Rev.\  D {\bf 75}, 054502 (2007)
  [arXiv:hep-lat/0610092].
%
\bibitem{ref:hyp:1}  A.~Hasenfratz and F.~Knechtli,
%\emph{Flavor symmetry and the static potential with hypercubic blocking}, 
Phys.~Rev.~D{\bf 64} (2002) 034504,
[{hep-lat/0103029}].
%
\bibitem{ref:asqtad:1}  K.~Orginos {\em et al.},
%\emph{Variants of fattening and flavor symmetry restoration}, 
Phys.~Rev.~D{\bf 60} (1999) 054503,
[{hep-lat/9903032}]; G.P.~Lepage, 
Phys.~Rev.~D{\bf 59} (1999) 074502,
[{hep-lat/9809157}].
%
\bibitem{ref:wlee:1} W.~Lee and S.R.~Sharpe, 
%\emph{Partial Flavor Symmetry Restoration for Chiral Staggered Fermions}, 
Phys.~Rev.~D{\bf 60} (1999) 094503,
[{hep-lat/9905023}].
%
%\bibitem{ref:wlee:2} Taegil Bae {\em et al.}, 
%%\emph{Pion spectrum using improved staggered fermions}, 
%PoS {\bf LAT2006} (2006) 086,
%[{hep-lat/0610057}].
%
\bibitem{ref:bernard:1} C.~Aubin and C.~Bernard, 
%\emph{Pion and kaon masses in staggered chiral perturbation theory}, 
Phys.~Rev.~D{\bf 68} (2003) 034014,
[{hep-lat/0304014}].
%
\bibitem{ref:SVdW} S.R.~Sharpe and R.S.~Van de Water,
% staggered chiral pert theory at NLO
Phys.~Rev.~D {\bf 71} (2005) 114505.
%
%\bibitem{ref:golterman:1} Maarten Golterman,
%%\emph{Staggered Mesons}, 
%Nucl.~Phys.~B{\bf 273} (1986) 663-676.
%
%\bibitem{ref:milc:2} C.~Bernard {\em et al.}, 
%%\emph{The QCD spectrum with three quark flavors.}, 
%Phys.~Rev.~D{\bf 64} (2001) 054506,
%[{hep-lat/0104002}].
%
\bibitem{ref:milc:1} C.~Aubin {\em et al.}, 
%\emph{Light pseudoscalar decay constants, quark masses, and
%low energy constants from three flavor lattice QCD}, 
Phys.~Rev.~D{\bf 70} (2004) 114501,
[{hep-lat/0407028}].
%



%
%\bibitem{ref:sharpe:1}  Ruth S.~Van de Water, Stephen R.~Sharpe, 
%%\emph{$B_K$ in staggered chiral perturbation theory}, 
%Phys.~Rev.~D{\bf 73} (2006) 014003,
%[{hep-lat/0507012}].
%
%\bibitem{ref:klu:1} H.~Kluberg-Stern {\em et al.},
%\emph{Flavors of Lagrangian Suskind fermions}, 
%Nucl.~Phys.~B{\bf 220} (1983) 447;
%D.~Verstegen, 
%\emph{Symmetry properties of fermionic bilinears $\cdots$},
%Nucl.~Phys.~B{\bf 249} (1985) 685.
%
%\bibitem{ref:ver:1} D.~Verstegen, 
%\emph{Symmetry properties of fermionic bilinears in lattice QCD},
%\emph{Nucl.~Phys.}  B{\bf 249} (1985) 685-703.
%
%\bibitem{ref:wlee:3} Weonjong Lee and Stephen Sharpe, 
%\emph{Perturbative matching of staggered four-fermion operators 
%with hypercubic fat links}, 
%Phys.~Rev.~D{\bf 68} (2003) 054510,
%[{hep-lat/0306016}].
%
\end{thebibliography}
\end{document}